\documentclass[12pt]{iopart}
\usepackage{iopams,graphicx}

\begin{document}

\review[Supernova relic neutrinos]{Relic neutrino background from
cosmological supernovae}
\author{Shin'ichiro Ando\dag\ and Katsuhiko Sato\dag\ddag}

\address{\dag\ Department of Physics, School of Science, The University
of Tokyo, 7-3-1 Hongo, Bunkyo-ku, Tokyo 113-0033, Japan}

\address{\ddag\ Research Center for the Early Universe, School of
Science, The University of Tokyo, 7-3-1 Hongo, Bunkyo-ku, Tokyo
113-0033, Japan}

\eads{\mailto{ando@utap.phys.s.u-tokyo.ac.jp},
\mailto{sato@phys.s.u-tokyo.ac.jp}}

\begin{abstract}
Present and future observations of supernova relic neutrinos (SRNs),
 i.e., a cosmological neutrino background from past core-collapse
 supernova explosions, potentially give us useful information concerning
 various fields of astrophysics, cosmology and particle physics.
We review recent progress of theoretical and observational studies of
 SRNs, particularly focusing on the detectability and also on
 implications for cosmic star formation history and neutrino physics.
\end{abstract}



\maketitle


\section{Introduction}
\label{sec:Introduction}

A core-collapse supernova explosion is one of the most spectacular
events in astrophysics, and it attracts a great deal of attention from
many physicists and astronomers.
It also produces a number of neutrinos and 99\% of its gravitational
binding energy is transformed to neutrinos; detection of the galactic
supernova neutrino burst by ground-based large water \v Cerenkov
detectors, such as Super-Kamiokande (SK) and Sudbury Neutrino
Observatory (SNO), would provide valuable information on the nature of
neutrinos as well as supernova physics.
In addition, because supernova explosions have occurred very commonly in
both the past and present universe, tracing the cosmic star formation
rate (SFR), they should have emitted a great number of neutrinos, which
now make a diffuse background, i.e., supernova relic neutrinos (SRNs).
Involved physics in SRN ranges quite widely---from cosmic SFR and
supernova physics to neutrino properties as elementary particles.
Therefore, detecting SRNs or even setting limits on their flux can give
us quite useful and unique implications for various fields of
astrophysics, cosmology and particle physics.
Detectability of SRNs in various detectors and its implications have
been discussed in many theoretical papers
\cite{Bisnovatyi-Kogan84}--\cite{Cocco04} from various points of view,
about which we discuss in detail in the following part of the present
paper.

Flux estimation requires models of neutrino spectrum emitted from each
supernova explosion and cosmic SFR.
Furthermore, it is experimentally established that neutrinos mix among
different flavours, altering the neutrino spectrum at detectors from the
original one after neutrinos propagate inside a supernova envelope; this
effect should now be taken into account appropriately.
In addition to a flux estimation as precise as possible, a detailed
discussion of background events, which hinder the SRN detection, is
essential; it has been thoroughly studied by Ando, Sato and Totani
\cite{Ando03a} and we follow their discussion later in this paper.
A stringent observational upper limit on the SRN flux is obtained by the
SK group \cite{Malek03} and its value is only a factor of 3--6 larger
than several theoretical predictions
\cite{Totani96a}--\cite{Hartmann97}, \cite{Ando03a}, being consistent
with (and more stringent than) the theoretical upper limit given by
Kaplinghat, Steigman and Walker \cite{Kaplinghat00} using conservative
models.
In order to make the SRN detection more likely, a promising method was
proposed by Beacom and Vagins \cite{Beacom04a}.
Their basic idea is to dissolve gadolinium trichloride (GdCl$_3$) into
the water \v Cerenkov detectors, which greatly reduces the background
events if it is applied to the currently working or proposed future
detectors such as SK, Hyper-Kamiokande (HK) and Underground Nucleon
Decay and Neutrino Observatory (UNO).
Therefore, we are now at an exciting stage, where SRNs would soon be
detected actually, and be used to obtain several implications for
various fields of astrophysics as a unique and complementary method to
usual observations of the light.

Since we have a promising prospects of the SRN detection, it is timely
to discuss the SRN potential for probing the universe or particle
physics.
Cosmic SFR is one of such possibilities.
The most popular method inferring SFR is to use the galaxy luminosity
function of rest-frame ultraviolet (UV) radiation
\cite{Lilly96}--\cite{Steidel99}, in addition to far-infrared
(FIR)/sub-millimeter dust emission \cite{Hughes98,Flores99} and
near-infrared (NIR) H$\alpha$ line emission
\cite{Gallego95}--\cite{Glazebrook99}.
In these traditional approaches, however, there are a fair number of
ambiguities when the actual observables are converted into the cosmic
SFR \cite{Somerville01}.
The most serious problem is that the effects of dust extinction are
non-negligible, especially for rest-frame UV observations.
For the approach using SRNs, on the other hand, we are not troubled by
such a problem, because neutrinos are completely free of dust
extinction.
This point is the same with observations in the sub-millimeter wave band;
however, neutrinos are emitted directly from stars, whereas
sub-millimeter radiation comes from dust and is an indirect process.
Another advantage is that supernovae are directly connected with the
death of massive stars with $M \gtrsim 8M_\odot$, whose lifetime is
expected to be very short compared with the Hubble time-scale
$H_0^{-1}$.
It enables direct inference of the cosmic SFR assuming the initial mass
function (IMF), not bothered with uncertainty concerning the galaxy
luminosity fuction, which sometimes causes difficulty in the case of
rest-frame UV observation \cite{Somerville01}.
Motivated by all of these reasons, many researchers have investigated
the dependence of the SRN flux on various models of cosmic SFR or to
what extent the SFR can be probed from the future SRN observations.
Totani, Sato and Yoshii \cite{Totani96a} gave flux estimation using
several theoretical models of galaxy evolution.
After the bursting release of the observational SFR data points since
the pioneering study by Madau \etal \cite{Madau96a}, the SRN flux has
been estimated using these SFR results
\cite{Malaney97,Hartmann97,Ando03a}.
More recently, theoretical SRN flux calculations have been compared with
the observational upper limit by SK \cite{Malek03}.
Fukugita and Kawasaki \cite{Fukugita03} used the limit to obtain
constraints on the cosmic SFR; Strigari \etal \cite{Strigari04} adopted
the latest SFR model based on the Sloan Digital Sky Survey
\cite{Glazebrook03} and concluded that their median SRN flux is slightly
below the current SK upper limit.
Following the proposal to dissolve Gd into detectors \cite{Beacom04a},
Ando \cite{Ando04a} investigated potential performance of the Gd-loaded
water \v Cerenkov detectors such as SK, HK and UNO, using the Monte
Carlo (MC) simulation, especially focusing on how reliably an assumed
SFR model can be reproduced from the SRN observation.

Neutrino properties can also be probed by the SRN observation in
principle.
It is pointed out that the resulting SRN signal depends on neutrino
oscillation models \cite{Ando03c}; particularly if the mass hierarchy is
inverted, i.e., the first mass eigenstate that most strongly couples to
electron flavour is heavier than the third state, $m_1>m_3$, and the
value of $\theta_{13}$, which has not been well constrained yet, is
sufficiently large, then the flavour conversion during propagation
inside the supernova envelope could considerably change the neutrino
signal.
Another interesting possibility is that a stringent constraint can be
set on neutrino decay models from the SRN observation as first pointed
out by Ando \cite{Ando03f} and succesively studied by Fogli \etal
\cite{Fogli04}.
Non-radiative neutrino decay can be induced by the interaction between
neutrinos and massless or very light particles such as Majoron.
The strongest lower limit to the neutrino lifetime-to-mass ratio
is obtained from the solar neutrino observation
\cite{Beacom02}--\cite{Eguchi03b} and meson decay
\cite{Barger82}--\cite{Britton94} to be $\tau /m \gtrsim 10^{-4}$
s/eV.
SRN observation, on the other hand, is sensitive to $\sim 10^{10}$ s/eV,
which is many orders of magnitude larger than the current limit.
This is because of the much longer baseline between the Earth and
cosmological supernovae, from which SRNs are emitted, than 1 AU in the
case of solar neutrino observations.
Neutrino decay could also change the detected signal from high-energy
astrophysical objects \cite{Beacom03} or the galactic supernova
explosions \cite{Frieman88}--\cite{Ando04c}, and could alter usual
discussions on the early universe and structure formation
\cite{Beacom04b} as well as on supernova coolings
\cite{Fuller88}--\cite{Farzan03}; but we stress that in principle
we can obtain the most stringent constraints on the neutrino lifetime
from the SRN observations.

This paper is organized as follows.
We first introduce formulation for calculating the SRN flux in
\sref{sec:Formulation and models}.
Models of cosmic SFR, supernova neutrino spectrum calculated by various
groups are given in the same section.
Neutrino oscillation inside the supernova envelope is briefly described
there.
SRN flux and event rate at ground-based detectors calculated with our
reference models are given in \sref{sec:Flux and event rate of supernova
relic neutrinos}.
\Sref{sec:Detectability and observational upper limit} is devoted to a
detailed discussion on background events against the SRN detection and
detectability at various detectors.
Current observational limit by SK, which is obtained by a statistical
argument including background events, is briefly summarized also in the
same section.
We discuss implications of the SRN observation for cosmic SFR and
neutrino properties such as oscillation and decay in sections
\ref{sec:Implication for cosmic star formation history} and
\ref{sec:Implication for neutrino properties}, respectively.
Finally, we conclude this paper by giving brief summary in
\sref{sec:Conclusions}.
Throughout this paper, we only consider electron anti-neutrinos
($\bar\nu_\rme$) at detectors because this kind of flavour is most
easily detected at the water \v Cerenkov detectors, on which we mainly
focus.

\section{Formulation and models}
\label{sec:Formulation and models}

\subsection{Formulation}
\label{sub:Formulation}

The present number density of SRNs ($\bar\nu_\rme$), whose energy is in
the interval $E_\nu\sim E_\nu +\rmd E_\nu$, emitted in the redshift
interval $z\sim z+\rmd z$, is given by
\begin{eqnarray}
 \rmd n_\nu(E_\nu)&=&R_{\rm SN}(z)(1+z)^3\frac{\rmd t}{\rmd z}
  \rmd z\frac{\rmd N_\nu (E_\nu^\prime)}{\rmd E_\nu^\prime}\rmd
  E_\nu^\prime (1+z)^{-3}\nonumber\\
 &=&R_{\rm SN}(z)\frac{\rmd t}{\rmd z}\rmd z\frac{\rmd N_\nu
  (E_\nu^\prime)}{\rmd E_\nu^\prime}(1+z)\rmd E_\nu ,
   \label{eq:dn_nu}
\end{eqnarray}
where $E_\nu^\prime =(1+z)E_\nu$ is the energy of neutrinos at redshift
$z$, which is now observed as $E_\nu$; $R_{\rm SN}(z)$ represents
the supernova rate per comoving volume at $z$, and hence the factor
$(1+z)^3$ should be multiplied to obtain the rate per physical volume at
that time; $\rmd N_\nu /\rmd E_\nu$ is the number spectrum of neutrinos
emitted by one supernova explosion; and the factor $(1+z)^{-3}$ comes
from the expansion of the universe.
The Friedmann equation gives the relation between $t$ and $z$ as
\begin{equation}
\frac{\rmd z}{\rmd t}=-H_0(1+z)\sqrt{\Omega_{\rm m}(1+z)^3
 +\Omega_\Lambda},
\label{eq:dz_dt}
\end{equation}
and we adopt the standard $\Lambda$CDM cosmology ($\Omega_{\rm m}=0.3,
\Omega_\Lambda =0.7,$ and $H_0=70~h_{70}~\mathrm{km~s^{-1}~
Mpc^{-1}}$).\footnote[3]{Although we use the specific cosmological model
here, the SRN flux itself is completely independent of such cosmological
parameters, as long as we use observationally inferred SFR models; see
their cancellation between equations \eref{eq:SRN flux} and
\eref{eq:SFR}.}
We now obtain the differential number flux of SRNs, $\rmd F_\nu /\rmd
E_\nu$, using the relation $\rmd F_\nu /\rmd E_\nu =c\,\rmd n_\nu /\rmd
E_\nu$:
\begin{equation}
 \frac{\rmd F_\nu}{\rmd E_\nu}=\frac{c}{H_0}\int_0^{z_{\rm max}}
  R_{\rm SN}(z)\frac{\rmd N_\nu (E_\nu^\prime)}{\rmd E_\nu^\prime}
 \frac{\rmd z}{\sqrt{\Omega_{\rm m}(1+z)^3+\Omega_\Lambda}},
  \label{eq:SRN flux}
\end{equation}
where we assume that gravitational collapses began at the redshift
$z_{\rm max}=5$.

\subsection{Models for cosmic star formation rate}
\label{sub:Models for cosmic star formation rate}

\begin{figure}[htbp]
\begin{center}
\includegraphics[width=8cm]{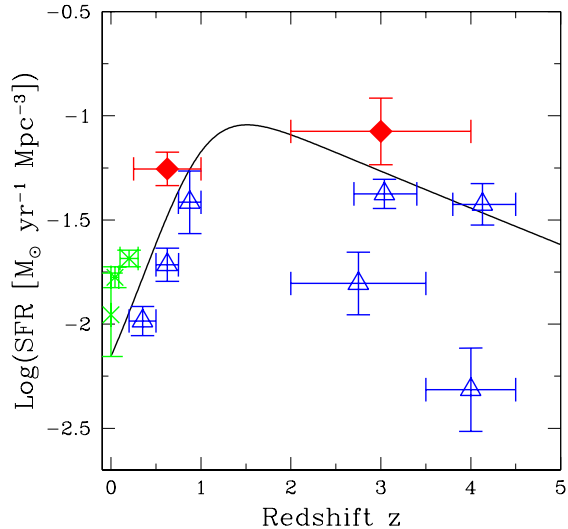}
\vspace{-5mm}
\caption{Cosmic star formation rate as a function of redshift. Data
 points are given by rest-frame UV ({\it open triangles})
 \cite{Lilly96,Madau96a,Steidel99}, NIR H$\alpha$ ({\it crosses})
 \cite{Gallego95,Gronwall98,Tresse98}, and FIR/sub-millimeter ({\it
 filled diamonds}) \cite{Hughes98,Flores99} observations. The solid
 curve represents our reference model given by equation \eref{eq:SFR}
 with $f_\ast = 1$. The standard $\Lambda$CDM cosmology is adopted
 ($\Omega_{\rm m}=0.3, \Omega_\Lambda =0.7, H_0=70$ km s${}^{-1}$
 Mpc${}^{-1}$) \label{fig:SFR}.}
\end{center}
\end{figure}

As our reference model for the SFR, we adopt a model that is based on
recent progressive results of rest-frame UV, NIR H$\alpha$, and
FIR/sub-millimeter observations; a simple functional form for the SFR per
unit comoving volume is given as \cite{Porciani01}
\begin{equation}
\fl
 \psi_\ast (z) = 0.32 f_\ast h_{70}
  \frac{\exp (3.4z)}{\exp (3.8z)+45}
 \frac{\sqrt{\Omega_{\rm m}(1+z)^3+\Omega_\Lambda}}{(1+z)^{3/2}}
 ~M_\odot~\mathrm{yr^{-1}~Mpc^{-3}},
  \label{eq:SFR}
\end{equation}
where $f_\ast$ is a factor of order unity, which we illustrate below.
\Fref{fig:SFR} shows the SFR $\psi_\ast (z)$ with the various data
points from rest-frame UV \cite{Lilly96,Madau96a,Steidel99}, H$\alpha$
line \cite{Gallego95}--\cite{Tresse98}, and FIR/sub-millimeter
\cite{Hughes98,Flores99} observations; these data points are not
corrected for dust extinction.
In the local universe, all studies show that the comoving SFR
monotonically increases with $z$ out to a redshift of at least 1.
Although there is such a general observational tendency, even the local
($z=0$) SFR density is far from precise determination; it ranges fairly
widely as $\psi_\ast (0)=$(0.5--2.9)$\times 10^{-2}h_{70}~M_\odot$
yr$^{-1}$ Mpc$^{-3}$ \cite{Baldry03}.
For this reason, we introduce the correction factor $f_\ast$.
Our reference model \eref{eq:SFR} with $f_\ast = 1$ is consistent
with mildly dust-corrected UV data at low redshift; on the other hand,
it may underestimate the results of the other wave band observations.
In fact, it predicts the local SFR value of $0.7\times 10^{-2} h_{70}
~M_\odot$ yr$^{-1}$ Mpc$^{-3}$, which is close to the lower limit of the
estimation by Baldry and Glazebrook \cite{Baldry03}, but we stress that
the SRN flux given below using this model can be simply applied to the
other cases if we adjust the correction factor $f_\ast$, and would be
quite general.
Calculations by Strigari \etal \cite{Strigari04} are based on the SFR
model giving local value of $1.6\times 10^{-2} ~M_\odot$ yr$^{-1}$
Mpc$^{-3}$, which corresponds to $f_\ast = 2.3$ in our notation.
Although the SFR-$z$ relation generally tends to increase from $z=0$ to
$\sim 1$, behaviours at the higher redshift region $z > 1$ are not clear
at all.
Ando \etal \cite{Ando03a} also investigated the dependence on the
several adopted SFR models, which were only different at high-redshift
regions ($z\gtrsim 1.5$); our reference model \eref{eq:SFR} was referred
to as the ``SF1'' model there.
They showed that the SRN flux at $E_\nu >10$ MeV is highly insensitive
to the difference among the SFR models (owing to the energy redshift, as
discussed in \sref{sub:Flux of supernova relic neutrinos}).

We obtain the supernova rate ($R_{\rm SN}(z)$) from the SFR by assuming
the Salpeter IMF ($\phi (m)\propto m^{-2.35}$) with a lower cutoff
around $0.5M_\odot$, and that all stars with $M>8M_\odot$ explode as
core-collapse supernovae, i.e.,
\begin{equation}
 R_{\rm SN}(z)=\frac{\int_{8M_\odot}^{125M_\odot}\rmd m~\phi(m)}
  {\int_{0}^{125M_\odot}\rmd m~m\phi(m)}\psi_\ast (z)
 =0.0122M_\odot^{-1}\psi_\ast (z).
  \label{eq:SN rate}
\end{equation}
Here we assume that the IMF does not change with time, which may be a
good approximation provided there are no significant correlations
between the IMF and the environment in which stars are born; extant
evidence seems to argue against such correlations over the redshift
range of interest ($z \lesssim 2$) \cite{Scalo98}.
The resulting local supernova rate evaluated with $f_\ast = 1$ agrees
within errors with the observed value of $R_{\rm SN}(0)=(1.2\pm
0.4)\times 10^{-4}h_{70}^3 \mathrm{~yr^{-1}~Mpc^{-3}}$ (e.g.,
\cite{Madau98b} and references therein).
In fact, the totally time-integrated neutrino spectrum from massive
stars ($\gtrsim 30M_\odot$) could be very different from the models that
we use (and give in the next subsection), possibly because of, e.g.,
black hole formation.
However, the conversion factor appearing in equation \eref{eq:SN rate}
is highly insensitive to the upper limit of the integral in the
numerator; for instance, if we change the upper limit in the numerator
to $25M_\odot$, the factor becomes $0.010M_\odot^{-1}$, which is only
slightly different from the value in equation \eref{eq:SN rate}.

\subsection{Neutrino spectrum from supernova explosions}
\label{sub:Neutrino spectrum from supernova explosions}

For the neutrino spectrum from each supernova, we adopt three reference
models by different groups, i.e., simulations by the Lawrence Livermore
(LL) group \cite{Totani98a} and Thompson, Burrows and Pinto
\cite{Thompson03} (hereafter TBP), and the MC study of spectral
formation by Keil, Raffelt and Janka \cite{Keil03} (hereafter KRJ).
In this field, however, the most serious problem is that the recent
sophisticated hydrodynamic simulations have not obtained the supernova
explosion itself; the shock wave cannot penetrate the entire core.
Therefore, many points still remain controversial, e.g., the average
energy ratio among neutrinos of different flavours, or how the
gravitational binding energy is distributed to each flavour.
All these problems are quite serious for our estimation, since the
binding energy released as $\bar\nu_\rme$ changes the normalization of
the SRN flux, and the average energy affects the SRN spectral shape.
Traditionally, neutrino spectrum is assumed to have a Fermi-Dirac
spectral shape with $T_{\bar\nu_\rme}\simeq 5$ MeV and $T_{\nu_{\rm
x}}\simeq 8$ MeV, where $\nu_{\rm x}$ represents non-electron neutrinos
and anti-neutrinos, as adopted in many studies including recent ones
\cite{Totani95}--\cite{Kaplinghat00},
\cite{Fukugita03,Strigari04,Cocco04}.
This approximation is roughly consistent with the LL model, although
slight difference exists at both high- and low-energy regions
\cite{Ando03a,Totani98a}.
The other two models (TBP and KRJ) that are more sophisticated, however,
are not consistent with such simple treatment at all as we describe
below or as already discussed in \cite{Ando04a}.
Because of all these reasons stated above, we believe that these three
models from different groups will be complementary.

The numerical simulation by the LL group \cite{Totani98a} is considered
to be the most appropriate for our estimation, because it is the only
model that succeeded in obtaining a robust explosion and in calculating
the neutrino spectrum during the entire burst ($\sim 15$ s).
According to their calculation, the average energy difference between
$\bar\nu_\rme$ and $\nu_{\rm x}$ was rather large and the complete
equipartition of the binding energy was realized $L_{\nu_\rme} =
L_{\bar\nu_\rme} = L_{\nu_{\rm x}}$, where $L_{\nu_\alpha}$ represents
the released gravitational energy as $\alpha$-flavour neutrinos.
The neutrino spectrum obtained by their simulation is well fitted by a
simple formula, which was originally given by KRJ as
\begin{equation}
 \frac{\rmd N_\nu}{\rmd E_\nu}=\frac{(1+\beta_\nu)^{1+\beta_\nu}L_\nu}
  {\Gamma (1+\beta_\nu)\bar E_\nu^2}
  \left(\frac{E_\nu}{\bar E_\nu}\right)^{\beta_\nu}
  e^{-(1+\beta_\nu)E_\nu/\bar E_\nu},
  \label{eq:beta fit}
\end{equation}
where $\bar E_\nu$ is the average energy; the values of the fitting
parameters for the $\bar\nu_\rme$ and $\nu_{\rm x}$ spectrum are
summarized in Table \ref{table:fitting parameters}.
\Table{\label{table:fitting parameters}Fitting parameters for supernova
neutrino spectrum.}
\br
 & Mass & $\bar E_{\bar\nu_\rme}$ & $\bar E_{\nu_{\rm x}}$ & & &
 $L_{\bar\nu_\rme}$ & $L_{\nu_{\rm x}}$ & \\
 Model & ($M_\odot$) & (MeV) & (MeV) & $\beta_{\bar\nu_\rme}$ & 
 $\beta_{\nu_{\rm x}}$ & (erg) & (erg)\\
\mr
LL \cite{Totani98a} & 20 & 15.4 & 21.6 & 3.8 & 1.8 & $4.9\times 10^{52}$
& $5.0\times 10^{52}$ \\
TBP \cite{Thompson03} & 11 & 11.4 & 14.1 & 3.7 & 2.2 & --- & ---\\
 & 15 & 11.4 & 14.1 & 3.7 & 2.2 & --- & ---\\
 & 20 & 11.9 & 14.4 & 3.6 & 2.2 & --- & --- \\
KRJ \cite{Keil03} & --- & 15.4 & 15.7 & 4.2 & 2.5 & --- & ---\\
\br
\endTable

Although the LL group succeeded in obtaining a robust explosion, their
result has recently been criticized because it lacked many relevant
neutrino processes that are now recognized as important.
Thus, we adopt the recent result of another hydrodynamic simulation, the
TBP one, which included all the relevant neutrino processes, such as
neutrino bremsstrahlung and neutrino-nucleon scattering with nucleon
recoil.
Their calculation obtained no explosion, and the neutrino spectrum ends
at 0.25 s after core bounce.
In the strict sense, we cannot use their result as our reference model
because the fully time-integrated neutrino spectrum is definitely
necessary in our estimate.
However, we adopt their result in order to confirm the effects of recent
sophisticated treatments of neutrino processes in the supernova core on
the SRN spectrum.
The TBP calculations include three progenitor mass models, i.e., 11, 15
and 20$M_\odot$; all of these models are well fitted by equation
\eref{eq:beta fit}, and the fitting parameters are summarized in
\tref{table:fitting parameters}.
The average energy for both $\bar\nu_\rme$ and $\nu_{\rm x}$ is much
smaller than that by the LL calculation.
Although we do not show this in \tref{table:fitting parameters}, it
was also found that at least for the early phase of the core-collapse,
the complete equipartition of the gravitational binding energy for each
flavour was not realized.
However, it is quite unknown whether these trends hold during the entire
burst.
In this study, we adopt the average energy given in
\tref{table:fitting parameters} as our reference model, while we assume
perfect equipartition between flavours, i.e.,
$L_{\bar\nu_\rme}=L_{\nu_{\rm x}}=5.0\times 10^{52}$ erg.

In addition, we also use the model by KRJ.
Their calculation did not couple with the hydrodynamics, but it focused
on the spectral formation of neutrinos of each flavour using an MC
simulation.
Therefore, the static model was assumed as a background of neutrino
radiation, and we use their ``accretion phase model II,'' in which the
neutrino transfer was solved in the background of a 150 ms postbounce
model by way of a general relativistic simulation.
The fitting parameters for their MC simulation are also summarized in
\tref{table:fitting parameters}.
Unlike the previous two calculations, their result clearly shows that
the average energy of $\nu_{\rm x}$ is very close to that of
$\bar\nu_\rme$.
It also indicates that the equipartition among each flavour was not
realized, but rather $L_{\nu_\rme}\simeq L_{\bar\nu_\rme}\simeq
2L_{\nu_{\rm x}}$.
However also in this case, since the totally time-integrated neutrino
flux is unknown from such temporary information, we assume perfect
equipartition, $L_{\bar\nu_\rme}=L_{\nu_{\rm x}}=5.0\times 10^{52}$
erg, as well as that the average energies are the same as those in
\tref{table:fitting parameters}.

\subsection{Neutrino spectrum after neutrino oscillation}
\label{sub:Neutrino spectrum after neutrino oscillation}

The original $\bar\nu_\rme$ spectrum is different from what we observe
as $\bar\nu_\rme$ at Earth, owing to the effect of neutrino
oscillation.
Since the specific flavour neutrinos are not mass eigenstates, they mix
with other flavour neutrinos during their propagation.
The behaviour of flavour conversion inside the supernova envelope is
well understood, because the relevant mixing angles and mass square
differences are fairly well determined by recent solar
\cite{Fukuda02a,Ahmed04}, atmospheric \cite{Fukuda99}, and reactor
neutrino experiments \cite{Eguchi03a}.
The remaining ambiguities concerning the neutrino oscillation parameters
are the value of $\theta_{13}$, which is only weakly constrained
($\sin^2\theta_{13}\lesssim 0.1$ \cite{Apollonio99}), and the type
of mass hierarchy, i.e., normal ($m_1\ll m_3$) or inverted ($m_1\gg
m_3$).
We first discuss the case of normal mass hierarchy as our standard
model; in this case, the value of $\theta_{13}$ is irrelevant.
The case of inverted mass hierarchy is addressed in
\sref{sub:Inverted mass hierarchy}.
In addition, other exotic mechanisms, such as resonant spin-flavour
conversion (see \cite{Ando03g} and references therein) and neutrino
decay \cite{Ando03f,Fogli04}, which possibly change the SRN flux and
spectrum, might work in reality, and these topics are also discussed
later in sections \ref{sub:Resonant spin-flavour conversion} and
\ref{sub:Decaying neutrinos}, respectively.

The produced $\bar\nu_\rme$ at the supernova core are coincident with
the lightest mass eigenstate $\bar\nu_1$ owing to the large matter
potentials.
Since this state $\bar\nu_1$ is the lightest also in vacuum, there are
no resonance regions in which one mass eigenstate can change into
another state, and therefore $\bar\nu_\rme$ at production arrives at the
stellar surface as $\bar\nu_1$.
Thus, the $\bar\nu_\rme$ spectrum observed by the distant detector is
\begin{eqnarray}
 \frac{\rmd N_{\bar\nu_\rme}}{\rmd E_{\bar\nu_\rme}}
  &=&|U_{\rme 1}|^2\frac{\rmd N_{\bar\nu_1}}{\rmd E_{\bar\nu_1}}
  +|U_{\rme 2}|^2\frac{\rmd N_{\bar\nu_2}}{\rmd E_{\bar\nu_2}}
  +|U_{\rme 3}|^2\frac{\rmd N_{\bar\nu_3}}{\rmd E_{\bar\nu_3}}\nonumber\\
  &=&|U_{\rme 1}|^2\frac{\rmd N_{\bar\nu_\rme}^0}{dE_{\bar\nu_\rme}}
   +\left(1-|U_{\rme 1}|^2\right)\frac{dN_{\nu_{\rm x}}^0}{dE_{\nu_{\rm x}}},
   \label{eq:spectrum after oscillation}
\end{eqnarray}
where the quantities with superscript $0$ represent those at production,
$U_{\alpha i}$ is the mixing matrix element between the
$\alpha$-flavour state and $i$-th mass eigenstate, and observationally
$|U_{\rme 1}|^2=0.7$.
In other words, 70\% of the original $\bar\nu_\rme$ survives; on the
other hand, the remaining 30\% comes from the other component $\nu_{\rm
x}$.
Therefore, both the original $\bar\nu_\rme$ and $\nu_{\rm x}$ spectra
are necessary for the estimation of the SRN flux and spectrum; since the
original $\nu_{\rm x}$ spectrum is generally harder than that of the
original $\bar\nu_\rme$, as shown in \tref{table:fitting parameters},
the flavour mixing is expected to harden the detected SRN spectrum.

\section{Flux and event rate of supernova relic neutrinos}
\label{sec:Flux and event rate of supernova relic neutrinos}

\subsection{Flux of supernova relic neutrinos}
\label{sub:Flux of supernova relic neutrinos}

\begin{figure}[htbp]
\begin{center}
\includegraphics[width=15cm]{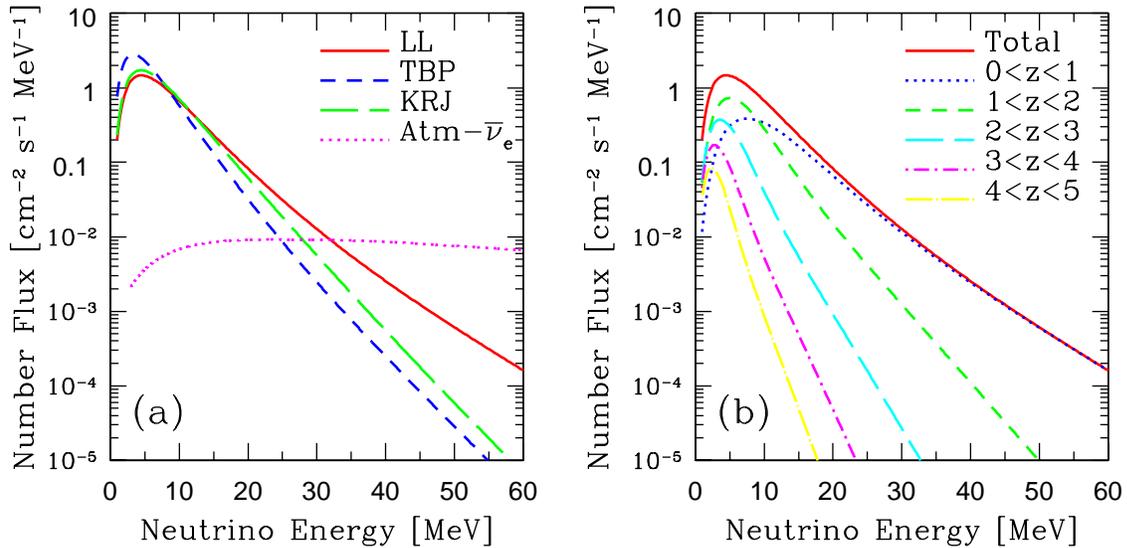}
\caption{(a) SRN flux in units of $f_\ast$ cm$^{-2}$ s$^{-1}$ MeV$^{-1}$
 calculated with three reference models of original neutrino spectrum:
 LL, TBP and KRJ. The flux of atmospheric neutrinos
 \cite{Gaisser88,Barr89} is also shown for comparison. (b) The same as
 (a), but indicating contribution from various redshift ranges. LL is
 adopted as the supernova model. These figures are taken from
 \cite{Ando04a}.\label{fig:SRN}}
\end{center}
\end{figure}

The SRN flux can be calculated by equation \eref{eq:SRN flux} with our
reference models given in \sref{sec:Formulation and models}.
\Fref{fig:SRN}(a) shows the SRN flux as a function of neutrino energy
for the three supernova models, LL, TBP and KRJ.
The flux of atmospheric $\bar\nu_\rme$, which becomes background events
for SRN detection, is shown in the same figure \cite{Gaisser88,Barr89}.
The SRN flux peaks at $\lesssim 5$ MeV, and around this peak, the TBP
model gives the largest SRN flux because the average energy of the
original $\bar\nu_\rme$ is considerably smaller than in the other two
models but the total released energy is assumed to be the same.
On the other hand, the model gives a smaller contribution at high-energy
regions, $E_\nu >10$ MeV.
In contrast, the high-energy tail of the SRN flux with the LL model
extends farther than with the other models, and it gives flux more than
a order of magnitude larger at $E_\nu =60$ MeV.
This is because the high-energy tail was mainly contributed by the
harder component of the original neutrino spectrum; in the case of the
LL calculation, the average energy of the harder component $\nu_{\rm x}$
is significantly larger than that of the other two calculations, as
shown in \tref{table:fitting parameters}.
We show the values of the SRN flux integrated over the various energy
ranges in \tref{table:flux}.
The total flux is expected to be 11--16 $f_\ast$ cm${}^{-2}$ s${}^{-1}$
for our reference models, although this value is quite sensitive to the
shape of the assumed SFR, especially at high-$z$.
The energy range in which we are more interested is high-energy regions
such as $E_\nu >19.3$ MeV and $E_\nu >11.3$ MeV, because as discussed
below, the background events are less critical and the reaction cross
section increases as $\propto E_\nu^2$.
In such a range, the SRN flux is found to be 1.3--2.3 $f_\ast$
cm${}^{-2}$ s${}^{-1}$ ($E_\nu >11.3$ MeV) and 0.14--0.46 $f_\ast$
cm${}^{-2}$ s${}^{-1}$ ($E_\nu >19.3$ MeV).
Thus, the uncertainty about the supernova neutrino spectrum and its
luminosity gives at least a factor 2--4 ambiguity to the expected SRN
flux in the energy region of our interest.
\Table{\label{table:flux}Flux of supernova relic neutrinos.}
\br
 & & \centre{3}{Flux ($f_\ast$ cm${}^{-2}$ s${}^{-1}$)}\\ \ns
 & & \crule{3}\\
 Model & Redshift range & Total & $E_\nu >11.3$ MeV & $E_\nu >19.3$ MeV\\
\mr
 LL & Total & 11.7 & 2.3 & 0.46\\
 & $0<z<1^{\rm a}$ & \04.1 (35.3) & 1.6 (70.9) & 0.39 (85.2)\\
 & $1<z<2^{\rm a}$ & \04.9 (42.0) & 0.6 (26.3) & 0.06 (14.0)\\
 & $2<z<3^{\rm a}$ & \01.8 (15.1) & 0.1 (\02.5) & 0.0\0 (\00.7)\\
 & $3<z<4^{\rm a}$ & \00.6 (\05.3) & 0.0 (\00.2) & 0.0\0 (\00.0)\\
 & $4<z<5^{\rm a}$ & \00.2 (\02.1) & 0.0 (\00.0) & 0.0\0 (\00.0)\\
 TBP & Total & 16.1 & 1.3 & 0.14\\
 KRJ & Total & 12.7 & 2.0 & 0.28\\
\br
\end{tabular}
\item[] $^{\rm a}$ Contribution from each redshift range to the total
 ($0<z<5$) value are shown in parentheses as percentages.
\end{indented}
\end{table}

\Fref{fig:SRN}(b) shows the contribution by supernova neutrinos emitted
from various redshift ranges.
At high-energy region $E_\nu >10$ MeV, the dominant flux comes from the
local supernovae ($0<z<1$), while the low-energy side is mainly
contributed by the high-redshift events ($z>1$).
This is because the energy of neutrinos that were emitted from a
supernova at redshift $z$ is reduced by a factor of $(1+z)^{-1}$
reflecting the expansion of the universe, and therefore high-redshift
supernovae only contribute to low-energy flux.
We also show the energy-integrated flux from each redshift range in
\tref{table:flux} in the case of the LL supernova model.
From the table, it is found that in the energy range of our interest,
more than 70\% of the flux comes from local supernova explosions at
$z<1$, while the high-redshift ($z>2$) supernova contribution is very
small.

\subsection{Event rate at water \v Cerenkov detectors}
\label{sub:Event rate at water Cerenkov detectors}

The water \v Cerenkov neutrino detectors have greatly succeeded in
probing the properties of neutrinos as elementary particles, such as
neutrino oscillation.
The SK detector is one of these detectors, and its large fiducial volume
(22.5 kton) might enable us to detect the diffuse background of SRNs.
Furthermore, much larger water \v Cerenkov detectors such as HK and UNO
are being planned.
SRN detection is most likely with the inverse $\beta$-decay reaction
with protons in water, $\bar\nu_\rme {\rm p}\to \rme^+{\rm n}$, and its
cross section is precisely understood \cite{Vogel99,Strumia03}.
In our calculation, we use the trigger threshold of SK-I (before the
accident).

\begin{figure}[htbp]
\begin{center}
\includegraphics[width=15cm]{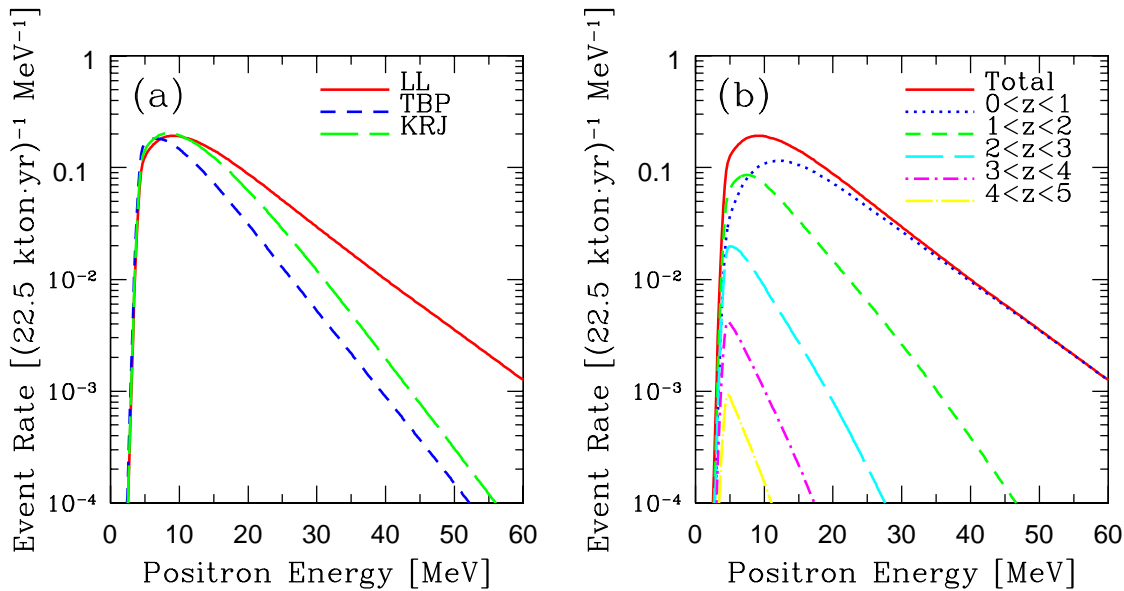}
\caption{(a) Event rate at water \v Cerenkov detectors in units of
 $f_\ast$ (22.5 kton yr)${}^{-1}$ MeV$^{-1}$ for three supernova
 models. (b) The same as (a), but indicating contribution from various
 redshift ranges. LL is adopted as the supernova model. These figures
 are taken from \cite{Ando04a}. \label{fig:evrt}}
\end{center}
\end{figure}

The expected event rates at such detectors are shown in
figures \ref{fig:evrt}(a) and \ref{fig:evrt}(b) in units of $f_\ast$
(22.5 kton yr)${}^{-1}$ MeV${}^{-1}$; with SK, it takes a year to obtain
the shown SRN spectrum, while with HK and UNO, much less time [$1~{\rm
yr}\times (22.5~{\rm kton}/V_{\rm fid})$, where $V_{\rm fid}$ is the
fiducial volume of HK or UNO] is necessary because of their larger
fiducial volume.
\Fref{fig:evrt}(a) compares the three models of the original supernova
neutrino spectrum, and \fref{fig:evrt}(b) shows the contribution to the
total event rate from each redshift range.
In \tref{table:event rate} we summarize the event rate integrated over
various energy ranges for three supernova models.
The expected event rate is 0.97--2.3 $f_\ast$ (22.5 kton yr)${}^{-1}$
for $E_\rme >10$ MeV and 0.25--1.0 $f_\ast$ (22.5 kton yr)${}^{-1}$ for
$E_\rme>18$ MeV.
This clearly indicates that if the background events that hinder the
detection are negligible, the SK has already reached the required
sensitivity for detecting SRNs; with the future HK and UNO, a
statistically significant discussion would be possible.
This also shows that the current shortage of our knowledge concerning the
original supernova neutrino spectrum and luminosity gives at least a
factor of 2 ($E_\nu >10$ MeV) to 4 ($E_\nu >18$ MeV) uncertainty to the
event rate at the high-energy range (actual detection range).
We also summarize the contribution from each redshift range in the same
table, especially for the calculation with the LL model.
The bulk of the detected events will come from the local universe
$(z<1)$, but the considerable flux is potentially attributed to the
range $1<z<2$.
\Table{\label{table:event rate}Event rate of supernova relic neutrinos.}
\br
 & & \centre{2}{Event rate [$f_\ast$ (22.5 kton yr)${}^{-1}$]} \\ \ns
 & & \crule{2}\\
 Model & Redshift range & $E_\rme >10$ MeV & $E_\rme >18$ MeV\\
\mr
 LL & Total & 2.3 & 1.0 \\
 & $0<z<1^{\rm a}$ & 1.7 (77.5) & 0.9 (87.5) \\
 & $1<z<2^{\rm a}$ & 0.5 (20.6) & 0.1 (11.9) \\
 & $2<z<3^{\rm a}$ & 0.0 (\01.7) & 0.0 (\00.5) \\
 & $3<z<4^{\rm a}$ & 0.0 (\00.1) & 0.0 (\00.0) \\
 & $4<z<5^{\rm a}$ & 0.0 (\00.0) & 0.0 (\00.0) \\
 TBP & Total & 0.97 & 0.25 \\
 KRJ & Total & 1.7 & 0.53 \\
\br
\end{tabular}
\item[] $^{\rm a}$ Contribution from each redshift range to the total
 ($0<z<5$) value are shown in parentheses as percentages.
\end{indented}
\end{table}

\section{Detectability and observational upper limit}
\label{sec:Detectability and observational upper limit}

\subsection{Background events against detection}
\label{sub:Background events against detection}

\begin{figure}[htbp]
\begin{center}
\includegraphics[width=15cm]{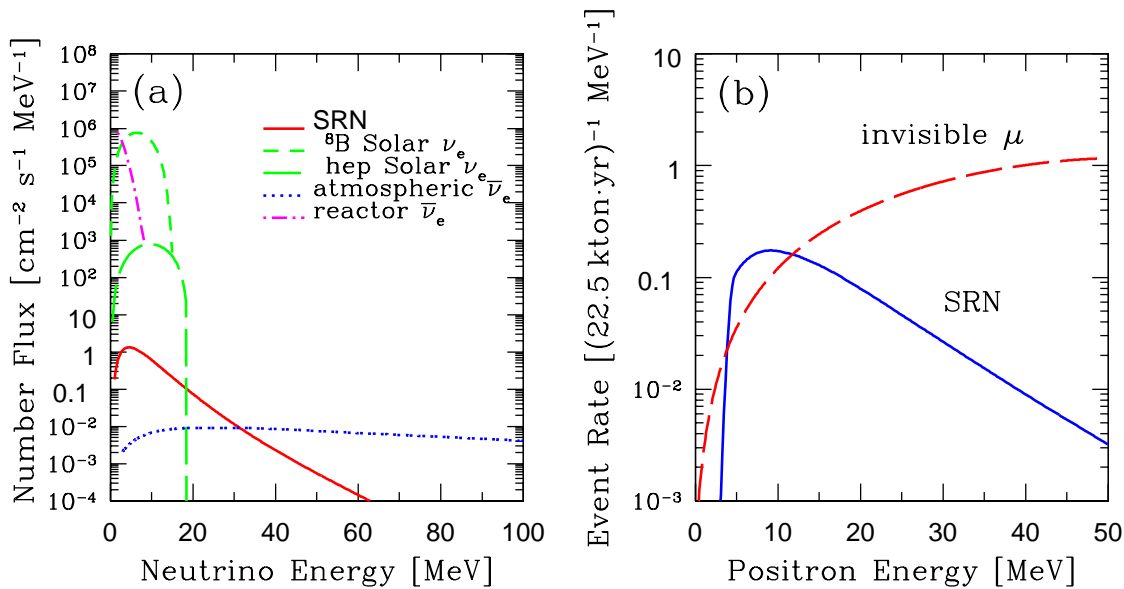}
 \caption{(a) Number flux of SRNs compared with other background
 neutrinos. (b) Event rate of SRNs and invisible muon decay products. In
 both panels, LL is adopted as the original neutrino spectrum. These
 figures are taken from \cite{Ando03a}. \label{fig:background}}
\end{center}
\end{figure}

In the previous section, we calculated the expected SRN spectrum at the
water \v Cerenkov detectors on the Earth, but the actual detection is
quite restricted because of the presence of other background events.
In this paper, we follow a detailed consideration of these backgrounds
by Ando \etal \cite{Ando03a}.
There are atmospheric and solar neutrinos, antineutrinos from nuclear
reactors, spallation products induced by cosmic-ray muons and decay
products of invisible muons.
We show in figures \ref{fig:background}(a) and \ref{fig:background}(b)
the flux and event rate of SRNs and these background events.

The flux of the atmospheric neutrinos is usually calculated using MC
method including various relevant effects (flux of primary cosmic rays,
solar modulation, geomagnetic field, interaction of cosmic rays in the
air, and so on), and in that simulation one-dimensional approximation is
used, i.e., after the interaction of primary cosmic ray particles with
air nuclei, all the particles are assumed to be moving along the line of
the momentum vector of the primary cosmic ray particles \cite{Kajita01}.
There are many authors who calculated the atmospheric neutrino flux
(see \cite{Lipari98a}--\cite{Honda95} for recent one-dimensional
results).
We use in this paper the flux calculated by Gaisser \etal
\cite{Gaisser88,Barr89}.
More recently, results of three-dimensional flux calculations have been
reported by several groups (see \cite{Barr04,Honda04} and references
therein).
These calculations show slight increase of the flux at low-energy
regions, although the flux below 100 MeV is not given.
Therefore, it should be noted that there is possibility that the flux of
atmospheric neutrinos in the relevant energy regime is higher than that
adopted in this paper by about 20--30\%.

Solar neutrino flux is dominant at energy range below 19 MeV.
We use the flux predicted by the standard solar model (SSM) in
\fref{fig:background} \cite{Bahcall89}.
Since the solar neutrinos are not $\bar\nu_\rme$ but $\nu_\rme$, the
cross section for them is about two orders of magnitude smaller than
that for $\bar\nu_\rme$.
Furthermore recoil electrons scattered by solar neutrinos strongly
concentrate to the opposite direction of the Sun, in contrast to the
isotropic distribution of $\bar\nu_\rme$ events.
Therefore, the solar neutrinos are an avoidable background, not as
critical as other events.
At the same energy range corresponding to $E_\nu \lesssim 19$ MeV, there
is another serious background that becomes an obstacle also for solar
neutrino detection, i.e., spallation products induced by cosmic ray
muons.
The event rate of the spallation background is several hundred per day
per 22.5 kton, and it is extremely difficult to reject all of these
events.
Therefore the solar neutrino range ($E_\nu\lesssim 19$ MeV), cannot be
used as an energy window.

The third background which we must consider is anti-neutrinos from
nuclear reactors.
In each nuclear reactor, almost all the power comes from the fissions of
the four isotopes, ${}^{235}$U ($\sim 75\%$), ${}^{238}$U ($\sim 7\%$),
${}^{239}$Pu ($\sim 15\%$) and ${}^{241}$Pu  ($\sim3\%$)
\cite{Bemporad02}.
Each isotope produces a unique electron anti-neutrino spectrum through
the decay of its fission fragments and their daughters.
The $\bar\nu_\rme$ spectrum from ${}^{235}$U, ${}^{239}$Pu and
${}^{241}$Pu can be derived using the semi-empirical formula with which
we fit data of detected $\beta$-spectrum from fission by thermal
neutrons \cite{Schreckenbach85,Hahn89}.
(${}^{238}$U undergoes only fast neutron fission and hence electron
spectrum from ${}^{238}$U cannot be measured by this kind of
experiment.)
Above 7 MeV, the number of $\beta$ counts drops dramatically and fitting
error becomes large.
In addition, with this method, as we determine the maximum $\beta$
energy and derives the energy distribution below that energy, it is
difficult to estimate the errors at the high-energy range.
While the $\bar\nu_\rme$ spectra in \cite{Schreckenbach85,Hahn89} are
given as tables, we use for simplicity somewhat less accurate analytical
approximation given in \cite{Vogel89}.
As a normalization factor we use energy-integrated $\bar\nu_\rme$ flux
at Kamioka, $1.34 \times 10^6$ cm$^{-2}$ s$^{-1}$, which are the
summation of the flux from various nuclear reactors in Japan and Korea
\cite{Bemporad02}.

With these backgrounds we discussed above and from
\fref{fig:background}(a), we expect the energy window of SRN events
ranging 19--30 MeV.
However, electrons or positrons from invisible muons are the largest
background in the energy window from 19 to 60 MeV.
This invisible muon event is illustrated as follows.
Atmospheric neutrinos produce muons by interaction with the nucleons
(both free and bound) in the fiducial volume.
If these muons are produced with energies below \v Cerenkov radiation
threshold (kinetic energy less than 53 MeV), then they will not be
detected (``invisible muons''), but successively produced electrons and
positrons from the muon decay will be visible.
Since this muon decay signal will mimic the $\bar\nu_\rme {\rm p} \to
\rme^+{\rm n}$ process in SK, it is difficult to distinguish SRN from
these events.
The spectrum of this invisible muon events is shown in
\fref{fig:background}(b), compared with the SRN spectrum.
Therefore, even at the remaining candidate of the energy window 19--30
MeV, there is a huge background due to the invisible muons, resulting in
{\it no energy window} for the SRN detection at present.

\subsection{Detectability at pure-water \v Cerenkov detectors}
\label{sub:Detectability at pure-water Cerenkov detectors}

\begin{figure}[htbp]
\begin{center}
\includegraphics[width=8cm]{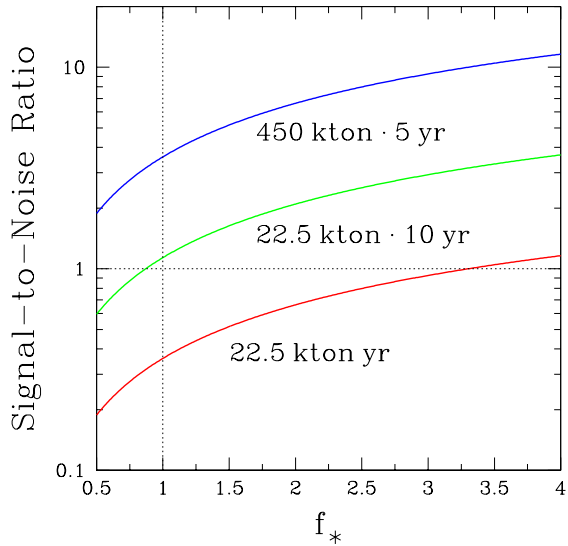}
\vspace{-5mm}
\caption{Signal-to-noise ratio $S/N$ of SRNs at pure-water \v Cerenkov
 detectors \eref{eq:SN ratio}, as a function of a correction factor
 $f_\ast$ for the SFR model \eref{eq:SFR}. LL is assumed for the
 original neutrino spectrum. Each line is labeled by the value of the
 effective volume $V_{\rm eff}$. \label{fig:SN_ratio}}
\end{center}
\end{figure}

We expect that SRNs can be most likely detected at water \v Cerenkov
detectors such as SK, because the largest class of volume can be
realized, enabling the most statistically significant discussions.
The most serious problem is that there is no energy window as we
described in the previous subsection.
In the energy range 19--30 MeV, the SRN event number is estimated to be
$N_{\rm SRN} = 0.73 f_\ast$($V_{\rm eff}/22.5 ~{\rm kton} \cdot{\rm
yr}$) for the LL model, where we define the effective volume of
detectors $V_{\rm eff}$ by (fiducial
volume)$\times$(time)$\times$(efficiency), which is the relevant
quantity representing the detector performance.
On the other hand, the background level would be as large as
$N_{\rm bg} \sim 3.4 (V_{\rm eff}/22.5 ~{\rm kton} \cdot {\rm yr})$,
mainly contributed by the invisible muon decay products.
Therefore, signal-to-noise ratio $S/N$ can be written as
\begin{equation}
S/N \equiv \frac{N_{\rm SRN}}{\sqrt{N_{\rm SRN} + N_{\rm bg}}}
 = \frac{0.73f_\ast}{\sqrt{0.73f_\ast + 3.4}}
 \left(\frac{V_{\rm eff}}{22.5 ~{\rm kton} ~{\rm yr}}\right)^{1/2},
\label{eq:SN ratio}
\end{equation}
plotted in \fref{fig:SN_ratio} as a function of $f_\ast$ for several
values of effective volume.
For detectors with the size of SK, one-year observation is found to be
insufficient to obtain significant SRN detection.
Further data-taking for about 10 years may reach $S/N \lesssim$ a few,
but this number is still insufficient in statistical significance.
Future mega-ton class detectors such as HK and UNO now planned
potentially give us a considerable number of SRN detection, by
data-taking for a couple of years, as can be seen from the upper curve
in \fref{fig:SN_ratio}.

\subsection{Detectability at other detectors}
\label{sub:Detectability at other detectors}

In this subsection, we discuss the SRN detectability at SNO (the heavy
water \v Cerenkov detector) and KamLAND (the scintillation detector),
which is mainly discussed in \cite{Kaplinghat00,Ando03a} and
\cite{Ando03a,Strigari04}, respectively.
An advantage of these detectors is that we are able to identify
$\bar\nu_\rme$ events using delayed coincidence signals; neutrons
produced via $\bar\nu_\rme\rmd\to\rme^+{\rm n}{\rm n}$ (SNO) and
$\bar\nu_\rme{\rm p}\to\rme^+{\rm n}$ (KamLAND) are tagged by deuterons,
Cl or He (SNO) and by protons (KamLAND), resulting in $\gamma$-ray
cascades, identified with the preceded positron signal.
Using this technique, we can remove other backgrounds from
non-$\bar\nu_\rme$ origin (solar neutrinos, invisible muons and
spallation products), opening up an energy window in the range of 10--30
MeV.
Unfortunately, because the detector volume is small (both detectors are
about 1 kton), the expected SRN rate is quite small, 0.03$f_\ast$
yr$^{-1}$ for SNO and 0.1$f_\ast$ yr$^{-1}$ for KamLAND \cite{Ando03a}.
More recently, Strigari \etal \cite{Strigari04} estimated the event rate
at KamLAND to be $\sim 0.4$ yr$^{-1}$, but they used the energy range
above 6 MeV, simply because there have been no events seen above this
energy at KamLAND \cite{Eguchi03a}.
In addition, the detectability at liquid argon detectors has been
discussed in \cite{Cocco04}; although the detection is still
challenging, potential advantage is that these detectors can mainly
capture $\nu_\rme$, which is difficult with other detectors.

\subsection{Current observational limit and future prospects}
\label{sub:Current observational limit and future prospects}

The most stringent upper limit on the SRN flux is obtained by the
observation for 1496 days (4.1 yr) at the SK detectors \cite{Malek03}.
This limit is obtained by the statistical analysis including the
background events from atmospheric neutrinos and invisible muons, and is
$< 1.2$ cm$^{-2}$ s$^{-1}$ over the energy region of $E_\nu >19.3$ MeV
(90\% CL).
Comparing with the prediction for the same energy range given in
\tref{table:flux}, we find that the current SK limit is only about a
factor 2.5--8.5 larger than our prediction using the reference model for
the cosmic SFR with $f_\ast = 1$, depending on the adopted original
neutrino spectrum.
This strong constraint motivated many theoretical studies
\cite{Fukugita03}--\cite{Strigari04} and has been translated
into constraints on various quantities as we further discuss in the
following sections.

Although the SK limit was derived by a careful analysis of the spectral
data over the energy range $E_\rme > 18$ MeV, we here show that a very
rough statistical argument using equation \eref{eq:SN ratio} can fairly
well reproduce the same limit.
In equation \eref{eq:SN ratio}, we consider setting a limit at 90\% CL,
which corresponds to $1.64\sigma$ level (or $S/N < 1.64$), as a result
of no SRN detection for $V_{\rm eff} = 22.5 ~{\rm kton}\cdot 4.1 ~{\rm
yr}$ observation.
Substituting these values into equation \eref{eq:SN ratio}, it is solved
for $f_\ast$, resulting in the upper limit $f_\ast < 2.6$ (90\% CL).
This value is then translated into the flux limit using the relation
between the SRN flux and $f_\ast$ (for the LL model) given in
\tref{table:flux}, i.e, $F_\nu < 1.2$ cm$^{-2}$ s$^{-1}$ for $E_\nu >
19.3 ~{\rm MeV}$ (90\% CL), which is in remarkable agreement with the
actual result reported by the SK group \cite{Malek03}.

In the near future, sensitivity of water \v Cerenkov detectors for the
SRN detection would be significantly improved by the promising technique
proposed recently \cite{Beacom04a}.
The basic idea is the same as the delayed coincidence technique actually
adopted by SNO or KamLAND (see discussions in \sref{sub:Detectability at
other detectors}), but GdCl$_3$ is dissolved into the pure-water of SK
(or other future detectors), which enables us to actively identify
$\bar\nu_\rme$ by capturing neutrons produced by the $\bar\nu_\rme {\rm
p}\to\rme^+{\rm n}$ reaction.
Owing to this proposal, the range 10--30 MeV would be an energy window
because we can positively distinguish the $\bar\nu_\rme$ signal from
other backgrounds such as solar neutrinos ($\nu_\rme$), invisible muon
events and spallation products.
The neutron capture efficiency by Gd is estimated to be 90\% with the
proposed 0.2\% admixture by mass of GdCl$_3$ in water, and subsequently
8 MeV $\gamma$-cascade occurs from the excited Gd.
The single-electron energy equivalent to this cascade was found to be
3--8 MeV by a careful simulation \cite{Hargrove95}, and with the trigger
threshold adopted in SK-I, only about 50\% of such cascades can be
detected actually.
However, it is expected that SK-III, which will begin operation in
mid-2006, will trigger at 100\% efficiency above 3 MeV, with good
trigger efficiency down to 2.5 MeV \cite{Beacom04a}.
In that case most of the $\gamma$-cascades from Gd will be detected with
their preceding signal of positrons.
From this point on, we assume 100\% efficiency; even if we abandon this
assumption, it does not affect our physical conclusion, since the
relevant quantity representing the detector performance is the effective
volume $V_{\rm eff}$ that already includes efficiency.

\section{Implication for cosmic star formation history}
\label{sec:Implication for cosmic star formation history}

\subsection{Constraints from the current observational limit}
\label{sec:Constraints from the current observational limit}

The current SK upper limit on the SRN flux is already stringent to give
some physical or astronomical consequences.
In this subsection, we discuss a constraint on the cosmic SFR from the
SRN limit; similar arguments have been given in \cite{Fukugita03}.
Comparing with our flux predictions using several supernova models
(\tref{table:flux}), the SK limit can be directly translated into the
bound on the correction parameter $f_\ast$ to the cosmic SFR introduced
in \eref{eq:SFR}, since the redshift dependence is roughly consistent
among various observations.
Further, the constraint on $f_\ast$ can then be used as that on the
local supernova rate and SFR.
The results are summarized in \tref{table:constraints} for various
supernova models.
For comparison, in the lower part of the same table, we also show the
result of local supernova surveys, $R_{\rm SN}(0)=(0.8$--$1.6)\times
10^{-4}$ yr$^{-1}$ Mpc$^{-3}$ \cite{Madau98b}, and the observationally
inferred SFR $\psi_\ast (0) = (0.5$--$2.9)\times 10^{-2}$ $M_\odot$
yr$^{-1}$ Mpc$^{-3}$ \cite{Baldry03}.
In particular for the local SFR, the SK limit on the SRN flux may rule
out some fraction of the observationally inferred value, if we choose
the LL model for the original neutrino spectrum; for the other two
supernova models, the SK limit is very close to the current upper bound
on the cosmic SFR by observations with the light.
Therefore, we stress that the neutrino observation has already reached
sensitivity to the cosmic SFR comparable with the usual and traditional
approaches using the light.
\Table{\label{table:constraints}Constraints on supernova rate and star
formation rate models. Upper part shows the SK limit on the SRN
observation \cite{Malek03}, whereas lower part shows the results of
local supernova surveys \cite{Madau98b} and observational inference of
the local SFR \cite{Baldry03}.}
\br
 & $f_\ast$ & $R_{\rm SN}(0)$ (yr$^{-1}$ Mpc$^{-3}$) & $\psi_\ast
(0)$ ($M_\odot$ yr$^{-1}$ Mpc$^{-3}$)\\
\mr
LL & $< 2.6$ & $< 2.2\times 10^{-4}$ & $< 1.8\times 10^{-2}$ \\
TBP & $< 8.6$ & $< 7.3\times 10^{-4}$ & $< 6.0\times 10^{-2}$ \\
KRJ & $< 4.3$ & $< 3.6\times 10^{-4}$ & $< 3.0\times 10^{-2}$ \\
\mr
\cite{Madau98b} & 0.9--1.9 & (0.8--1.6)$\times 10^{-4}$ & --- \\
\cite{Baldry03} & 0.7--4.2 & --- & (0.5--2.9)$\times 10^{-2}$ \\
\br
\endTable

\subsection{Performance of Gd-loaded detectors}
\label{sub:Performance of Gd-loaded detectors}

Performance of the proposed Gd-loaded detectors as an SFR probe is of
our interest, and has recently been investigated in detail by Ando
\cite{Ando04a} using the MC simulation; in this subsection, we briefly
introduce his discussion.
Although we focus here on how far the SFR can be probed by SRN
observation, the uncertainty from the supernova neutrino spectrum would
give a fair amount of error.
However, this problem can be solved if a supernova explosion occurs in
our galaxy; the expected event number is about 5000--10,000 at SK, when
supernova neutrino burst occurs at 10 kpc, and it will enable a
statistically significant discussion concerning the neutrino spectrum
from supernova explosions.
Even if there are no galactic supernovae in the near future, remarkable
development of the supernova simulation can be expected with the growth
of computational resources and numerical technique.
With such developments, the supernova neutrino spectrum and luminosity
may be uncovered, and the ambiguity is expected to be reduced
significantly.
Thus, in this paper we assume that the supernova neutrino spectrum is
well understood and that our reference models are fairly good
representatives of nature; we analyze the SFR alone with several free
parameters.

The basic procedure adopted in \cite{Ando04a} is as follows.
(1) The expected signal (spectrum) at a Gd-loaded detector in the range
10--30 MeV is simulated, assuming that there are no background events.
In that process, the SFR given by equation \eref{eq:SFR} with $f_\ast =
1$ and the LL model as neutrino spectrum are used for the generation of
the SRN signal.
(2) Then the SRN spectrum is analyzed using the maximum likelihood
method with two free parameters of the SFR and a set of the best-fit
values for those parameters is obtained; they are concerned with the
supernova rate as
\begin{equation}
R_{\rm SN}(z)=\left\{
	       \begin{array}{ccc}
		R_{\rm SN}^0 (1+z)^\alpha & {\rm for} & z<1,\\
		2^\alpha R_{\rm SN}^0 & {\rm for} & z>1,
\end{array}\right.
\label{eq:parameterization}
\end{equation}
where $R_{\rm SN}^0$ represents the local supernova rate and $\alpha$
determines the slope of supernova rate evolution.
Although the constant SFR at $z>1$ is assumed, it is found that the
result would be the same even if this assumption is changed.
This is because the bulk of the detected event comes from local
supernova as already shown.
(3) 10$^3$ such MC simulations are performed and 10$^3$ independent sets
of best-fit parameters are obtained.
Then we discuss the standard deviation of the distributions of such
best-fit parameter sets and the implications for the cosmic SFR.

\begin{figure}[htbp]
\begin{center}
\includegraphics[width=15cm]{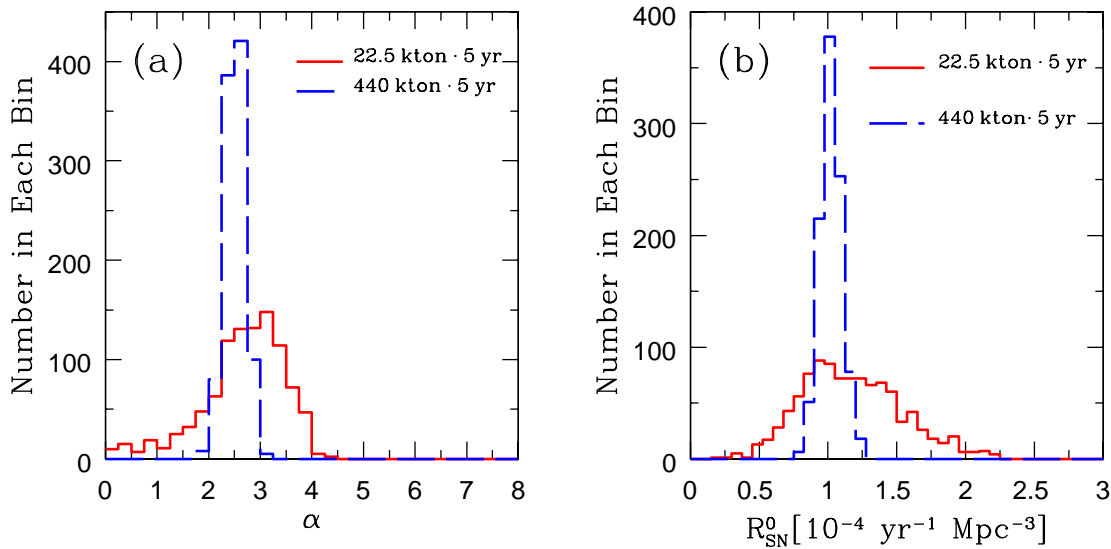}
\caption{Distribution of 10$^3$ best-fit values for (a) $\alpha$ and (b)
 $R_{\rm SN}^0$, which are obtained from the analyses of each MC
 generation. The effective volume is $22.5 {\rm ~kton}\cdot 5 ~{\rm yr}$
 for solid histogram and $440 ~{\rm kton}\cdot 5 ~{\rm yr}$ for dashed
 histogram. (a) The value of the local supernova rate is fixed to be
 $R_{\rm SN}^0=1.2\times 10^{-4}$ yr${}^{-1}$ Mpc${}^{-3}$. The
 resulting distributions are characterized by $\alpha = 2.7\pm 0.8$ for
 an effective volume of $22.5 ~{\rm kton}\cdot 5 ~{\rm yr}$ and $\alpha
 = 2.5\pm 0.2$ for $440 ~{\rm kton}\cdot 5 ~{\rm yr}$. (b) The value of
 $\alpha$ is fixed to be 2.9. The resulting distributions are
 characterized by $R_{\rm SN}^0 = (1.2\pm 0.4)\times 10^{-4}$ yr$^{-1}$
 Mpc$^{-3}$ for $22.5 ~{\rm kton}\cdot 5 ~{\rm yr}$ and $R_{\rm SN}^0 =
 (1.0\pm 0.1)\times 10^{-4}$ yr$^{-1}$ Mpc$^{-3}$ for $440 ~{\rm kton}
 \cdot 5 ~{\rm yr}$. These figures are taken from
 \cite{Ando04a}. \label{fig:MC_hist}}
\end{center}
\end{figure}

\begin{figure}[htbp]
\begin{center}
\includegraphics[width=8cm]{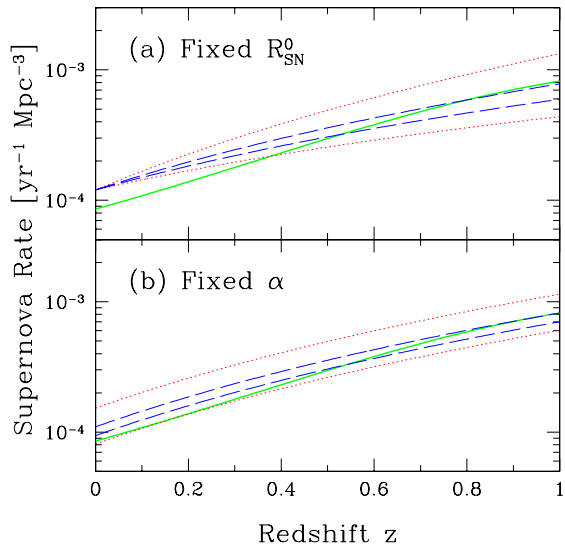}
\vspace{-3mm}
\caption{Supernova rate as a function of redshift. In both panels, solid
 curves represent the reference model, from which MC data were
 generated. (a) The allowed region at the $1\sigma$ level, concerning
 the fitting parameter $\alpha$ with fixed $R_{\rm SN}^0$, is shown as
 the area between the two dotted curves for an effective volume of $22.5
 ~{\rm kton}\cdot 5 ~{\rm yr}$ and as the area between the two dashed
 curves for an effective volume of $440 ~{\rm kton} \cdot 5 ~{\rm yr}$.
 (b) The same as (a) but for fitting parameter $R_{\rm SN}^0$ with fixed
 $\alpha$. This figure is taken from \cite{Ando04a}. \label{fig:SNrate}}
\end{center}
\end{figure}

First we discuss the performance of Gd-SK for 5 years, or an effective
volume of $22.5 ~{\rm kton}\cdot 5 ~{\rm yr}$.
Because the expected event number is only $\sim 10$, the parameters
$R_{\rm SN}^0$ and $\alpha$ cannot both be well determined at once.
Therefore, one of those parameters should be fixed in advance at some
value inferred from other observations.
First, the value of $R_{\rm SN}^0$ was fixed to be $1.2\times 10^{-4}$
yr${}^{-1}$ Mpc${}^{-1}$, which was inferred from the local supernova
survey \cite{Madau98b}, and the distribution of the best-fit values of
parameter $\alpha$ was obtained.
The result of 10$^3$ MC simulations are shown in \fref{fig:MC_hist}(a)
as a histogram of the distribution of best-fit parameters $\alpha$ ({\it
solid histogram}).
The average value of these 10$^3$ values for $\alpha$ is found to be
2.7, and the standard deviation is 0.8, i.e., $\alpha =2.7\pm 0.8$.
Then in turn, the value of $\alpha$ was fixed to be 2.9 in order to
obtain the distribution of best-fit values for the local supernova rate
$R_{\rm SN}^0$ from the SRN observation.
The result of 10$^3$ MC generations and analyses in this case is shown in
\fref{fig:MC_hist}(b).
The average value for $R_{\rm SN}^0$ is $1.2\times 10^{-4}$ yr${}^{-1}$
Mpc${}^{-3}$, and the standard deviation is $0.4\times 10^{-4}$
yr${}^{-1}$ Mpc${}^{-3}$, i.e., $R_{\rm SN}^0=(1.2\pm 0.4)\times
10^{-4}$ yr${}^{-1}$ Mpc${}^{-3}$.
\Fref{fig:SNrate} shows the comparison between the supernova rate model,
in which the parameter is inferred from the MC simulations, and the
``true'' reference model; the cases of fixed $R_{\rm SN}^0$ and $\alpha$
are shown in figures \ref{fig:SNrate}(a) and \ref{fig:SNrate}(b),
respectively.
The allowed region at the $1\sigma$ level is located between the two
dotted curves, while the solid curve represents the reference model,
from which MC data were generated.
Thus, with the Gd-SK detector we can roughly reproduce the supernova
rate profile at $z<1$ for 5 years operation.

Now we turn our attention to future mega-ton class detectors such as
Gd-HK or Gd-UNO.
With these detectors, the effective volume considered, $440 ~{\rm
kton}\cdot 5 ~{\rm yr}$, is expected to be realized in several years
from the start of their operation.
First the same analysis as illustrated above was performed; i.e., one of
relevant parameters, $\alpha$ or $R_{\rm SN}^0$, was fixed and the
dependence on the remaining parameter was investigated.
The result of these cases are also shown in figures \ref{fig:MC_hist}(a)
and \ref{fig:MC_hist}(b) as dashed histograms, which give $\alpha
=2.5\pm 0.2$ and $R_{\rm SN}^0=(1.0\pm 0.1)\times 10^{-4}$ yr${}^{-1}$
Mpc${}^{-3}$, respectively.
The statistical errors are considerably reduced compared with the case
of $22.5 ~{\rm kton} \cdot 5 ~{\rm yr}$, because of the $\sim 20$ times
larger effective volume.
Thus, future maga-ton detectors will possibly pin down, within 10\%
statistical error, either the index of supernova rate evolution $\alpha$
or the local supernova rate $R_{\rm SN}^0$ if the other is known in
advance.
The dashed curves in \fref{fig:SNrate} set the allowed region of
the supernova rate at the $1\sigma$ level by the considered detectors,
well reproducing the assumed model.
In principle, we can determine both parameters by SRN observation,
because $R_{\rm SN}^0$ is concerned with the absolute value of the flux
alone but $\alpha$ is concerned with both the absolute value and the
spectral shape; i.e., these two parameters are not degenerate with each
other.
Thus, the same procedure was repeated but without fixing the values of
$\alpha$ or $R_{\rm SN}^0$.
The distribution of 10$^3$ best-fit parameter sets of ($\alpha,R_{\rm
SN}^0$) was calculated assuming a detector with an effective volume of
$440 ~{\rm kton}\cdot 5 ~{\rm yr}$; the mean values and the standard
deviations were found to be $\alpha =3.5\pm 1.3$ and $R_{\rm
SN}^0=(8.8\pm 4.8)\times 10^{-5}$ yr${}^{-1}$ Mpc${}^{-3}$.
Even though the effective volume is as large as $440 ~{\rm kton}\cdot 5
~{\rm yr}$, it is still insufficient for determining both parameters at
once.

Throughout the above arguments or those in \cite{Ando04a}, the reference
SFR model with $f_\ast = 1$ was assumed and they seem to be dependent on
the adopted values of $f_\ast$.
However, the results can easily be applied to the case with other values
of $f_\ast$.
This is because the statistical errors are inversely proportinal to the
square root of the event number, which is simply proportional to $f_\ast
V_{\rm eff}$ in our arguments; i.e., $\delta\alpha / \langle \alpha
\rangle, \delta R_{\rm SN}^0 / \langle R_{\rm SN}^0 \rangle \propto
(f_\ast V_{\rm eff})^{-1/2}$.
In consequence, we find that the discussions given above and in
\cite{Ando04a} are quite general and can be easily modified in the other
cases of $f_\ast$ using the following relations:
\begin{equation}
\fl
 \frac{\delta \alpha}{\langle \alpha \rangle}
  \simeq 0.3 \left(\frac{f_\ast V_{\rm eff}}
	   {22.5 ~{\rm kton}\cdot 5 ~{\rm yr}}\right)^{-1/2}, ~~~
 \frac{\delta R_{\rm SN}^0}{\langle R_{\rm SN}^0 \rangle}
  \simeq 0.3 \left(\frac{f_\ast V_{\rm eff}}
	   {22.5 ~{\rm kton}\cdot 5 ~{\rm yr}}\right)^{-1/2}.
  \label{eq:errors}
\end{equation}

As a next step, it would be useful to investigate the possibility that
how far (to which $z$) we can probe the cosmic SFR by the SRN
observations.
As we have already shown, the main contribution to the SRN even rate at
10--30 MeV comes from low-redshift region $0<z<1$.
Signals from further high-redshift universe would become enhanced if we
could reduce the lower energy threshold $E_{\rm th}$ of the SRN
detection range.
The value of $E_{\rm th}$ is restricted to 10 MeV because at energy
regions lower than this, there is a large background of reactor
neutrinos; its removal is impossible with the current detection
methods.
Since the SK and HK detectors are (will be) located at Kamioka in Japan,
they are seriously affected by background neutrinos from many nuclear
reactors.
If some large-volume detectors were built at a location free from such
background, the lower threshold energy could be reduced, enabling us
to probe the high-redshift supernova rate.
Ando \cite{Ando04a} also discussed performance of future mega-ton class
detectors as a function of the value of $E_{\rm th}$ and found that the
behaviour of SFR at $1<z<2$ more and more affects the detected event
number at $E_{\rm th} < E_\rme < 30$ MeV, as we reduce the threshold
energy.

\section{Implication for neutrino properties}
\label{sec:Implication for neutrino properties}

\subsection{Inverted mass hierarchy}
\label{sub:Inverted mass hierarchy}

Throughout the above discussions, we have assumed normal hierarchy of
neutrino masses ($m_1\ll m_3$).
However, the case of inverted mass hierarchy has not been experimentally
excluded yet, and we explore this possibility in this subsection
following the discussion given by Ando and Sato \cite{Ando03c}.
In this case, flavour conversions inside the supernova envelope change
dramatically, compared with the normal mass hierarchy already discussed
in \sref{sub:Neutrino spectrum after neutrino oscillation}.
Since $\bar\nu_3$ is the lightest, $\bar\nu_\rme$ are created as
$\bar\nu_3$, owing to large matter potential.
In that case, it is well known that at a so-called resonance point,
there occurs a level crossing between $\bar\nu_1$ and $\bar\nu_3$ (for a
more detailed discussion, see, e.g., \cite{Dighe00}).
At this resonance point, complete $\bar\nu_1\leftrightarrow\bar\nu_3$
conversion occurs when the so-called adiabaticity parameter is
sufficiently small compared to unity (it is said that resonance is
``non-adiabatic''), while conversion never occurs when it is large
(adiabatic resonance).
The adiabaticity parameter $\gamma$ is quite sensitive to the value of
$\theta_{13}$, i.e., $\gamma\propto\sin^22\theta_{13}$; when
$\sin^22\theta_{13}\gtrsim 10^{-3}$ ($\sin^22\theta_{13}\lesssim
10^{-5}$), the resonance is known to be completely adiabatic
(non-adiabatic) \cite{Dighe00}.
When the resonance is completely non-adiabatic (because of small
$\theta_{13}$), the situation is the same as in the case of normal mass
hierarchy already discussed in \sref{sub:Neutrino spectrum after
neutrino oscillation} (because $\bar\nu_\rme$ at production become
$\bar\nu_1$ at the stellar surface), and the $\bar\nu_\rme$ spectrum
after oscillation is represented by equation (\ref{eq:spectrum after
oscillation}).
On the other hand, adiabatic resonance (due to large $\theta_{13}$)
forces $\bar\nu_\rme$ at production to become $\bar\nu_3$ when they
escape from the stellar surface, and therefore the observed
$\bar\nu_\rme$ spectrum is given by
\begin{equation}
 \frac{\rmd N_{\bar\nu_\rme}}{\rmd E_{\bar\nu_\rme}}
  =|U_{\rme 3}|^2\frac{\rmd N_{\bar\nu_\rme}^0}{\rmd E_{\bar\nu_\rme}}
   +\left(1-|U_{\rme 3}|^2\right)\frac{\rmd N_{\nu_{\rm x}}^0}{\rmd
   E_{\nu_{\rm x}}}
   \simeq\frac{\rmd N_{\nu_{\rm x}}^0}{\rmd E_{\nu_{\rm x}}}.
   \label{eq:spectrum after oscillation for inverted hierarchy}
\end{equation}
The second equality follows from the fact that the value of $|U_{\rme
3}|^2$ is constrained to be much smaller than unity from reactor
experiments \cite{Apollonio99}.
Thus, equation \eref{eq:spectrum after oscillation for inverted
hierarchy} indicates that complete conversion takes place between
$\bar\nu_\rme$ and $\nu_{\rm x}$.
When the value of $\theta_{13}$ is large enough to induce adiabatic
resonance ($\sin^22\theta_{13}\gtrsim 10^{-3}$), the obtained SRN flux
and spectrum should be very different from ones obtained in
sections \ref{sub:Flux of supernova relic neutrinos} and \ref{sub:Event
rate at water Cerenkov detectors}.
The SRN flux and event rate in this case were calculated with equations
\eref{eq:SRN flux} and \eref{eq:spectrum after oscillation for inverted
hierarchy}, and the results are summarized in tables \ref{table:flux for
inverted hierarchy} and \ref{table:event rate for inverted hierarchy}.
The values (with the LL model) shown in this table are consistent with
the calculation by \cite{Ando03c}, in which numerically calculated
conversion probabilities were adopted with some specific oscillation
parameter sets (which include a model with inverted mass hierarchy and
$\sin^22\theta_{13}=0.04$), as well as realistic stellar density
profiles; they also included the shock wave propagation and the Earth
matter effect in their calculations, but both were found to affect only
by a few percent and irrelevant.
\Table{\label{table:flux for inverted hierarchy}Flux of supernova relic
neutrinos in the case of inverted mass hierarchy. Values given in this
table are applicable only when the value of $\theta_{13}$ is large
enough to induce completely adiabatic resonance, i.e., $\sin^2
2\theta_{13} \gtrsim 10^{-3}$. If $\sin^2 2\theta_{13} \lesssim
10^{-5}$, on the other hand, the results become the same as those given
in \tref{table:flux}.}
\br
 & \centre{3}{Flux ($f_\ast$ cm${}^{-2}$ s${}^{-1}$)}\\ \ns
 & \crule{3}\\
 Model & Total & $E_\nu >11.3$ MeV & $E_\nu >19.3$ MeV\\
\mr
 LL & \09.4 & 3.1 & 0.94\\
TBP & 13.8 & 1.9 & 0.30\\
KRJ & 12.7 & 2.2 & 0.38\\
\br
\endTable
\Table{\label{table:event rate for inverted hierarchy}Event rate of
supernova relic neutrinos in the case of inverted mass hierarchy. Values
given in this table are applicable only when the value of $\theta_{13}$
is large enough to induce completely adiabatic resonance, i.e., $\sin^2
2\theta_{13} \gtrsim 10^{-3}$. If $\sin^2 2\theta_{13} \lesssim
10^{-5}$, on the other hand, the results become the same as those in
\tref{table:event rate}.}
\br
 & \centre{2}{Event rate [$f_\ast$ (22.5 kton yr)${}^{-1}$]} \\ \ns
 & \crule{2}\\
 Model & $E_\rme >10$ MeV & $E_\rme >18$ MeV\\
\mr
 LL & 3.8 & 2.3 \\
 TBP & 1.6 & 0.58 \\
 KRJ & 2.0 & 0.76 \\
\br
\endTable

The total flux becomes 9.4--14 $f_\ast$ cm${}^{-2}$ s${}^{-1}$, somewhat
smaller than the values given in \tref{table:flux}, because the total
flux is dominated by the low-energy region.
The fluxes at $E_\nu >19.3$ MeV are enhanced to be 0.30--0.94 $f_\ast$
cm${}^{-2}$ s${}^{-1}$, but this is still below the current 90\% CL
upper limit of $1.2$ cm${}^{-2}$ s${}^{-1}$ obtained by the SK
observation if $f_\ast = 1$ is adopted.
The event rate at the future detectable energy range, $E_\nu >10$ MeV,
is expected to become 1.6--3.8 $f_\ast(22.5 ~{\rm kton ~yr})^{-1}$,
which is considerably larger than the values in the case of normal mass
hierarchy, 0.97--2.3 $f_\ast (22.5 ~{\rm kton ~yr})^{-1}$.
The increase (decrease) of the flux and event rate integrated over the
high (total) energy region is due to the very high efficiency of the
flavour conversion, $\nu_{\rm x}\to\bar\nu_\rme$, inside the supernova
envelope; because the original $\nu_{\rm x}$ are expected to be produced
with larger average energy as shown in \tref{table:fitting parameters},
the efficient conversion makes the SRN spectrum harder, which enhances
the flux and event rate at the high-energy region.
Thus, if the inverted mass hierarchy, as well as the large value for
$\theta_{13}$, were realized in nature, SRN detection would be rather
easier, compared with the other cases.
Although we do not repeat the MC simulations that were introduced in
\sref{sub:Performance of Gd-loaded detectors}, the results can be easily
inferred; the statistical errors in this case would be $\sim
(3.8/2.3)^{1/2}=1.3$ times smaller than the values given in equation
\eref{eq:errors}, because they are inversely proportional to the square
root of the event number.

\subsection{Resonant spin-flavour conversion}
\label{sub:Resonant spin-flavour conversion}

If neutrinos have non-zero magnetic moment as large as $10^{-12}\mu_{\rm
B}$, where $\mu_{\rm B}$ is the Bohr magneton, it potentially changes
supernova neutrino signal owing to an additional effect of the flavour
conversions.
Especially if neutrinos are the Majorana particles, the interaction
between the Majorana magnetic moment and supernova magnetic fields
induces a spin-flavour conversion (e.g., $\bar\nu_\rme \leftrightarrow
\nu_{\mu,\tau}, \nu_\rme \leftrightarrow \bar\nu_{\mu,\tau}$) resonantly
(see \cite{Ando03g} and references therein).
This mechanism could potentially give quite a characteristic supernova
neutrino signal at detectors, and also is expected to affect the SRN
spectrum significantly.
However, it is still premature to estimate the SRN flux including the
spin-flavour conversions for several reasons.
First, the shock wave propagation can change the magnetic field
structure as well as the density profile of supernova progenitors, both
of which are essential in calculating flavour-conversion probabilities
\cite{Ando03b}.
Because there is no reliable supernova simulation that succeeded in
pushing the shock wave outside the core, we are even not at the stage to
start the calculation.
Furthermore, calculating how the propagating shock wave changes the
magnetic field structure would be quite a difficult task.
The second reason is that the spin-flavour conversion probabilities
strongly depend on the metallicity of progenitor stars \cite{Ando03d}.
Because SRN is the accumulation of neutrinos from past supernova
explosions, it should be quite natural that the poor metal stars give
some contribution to the SRN flux.
In consequence, it would be difficult to obtain some implication for the
neutrino magnetic moment or supernova magnetic fields from the SRN
detection at present; instead, a future galactic supernova neutrino
burst might give some clues by its time profile or spectrum
\cite{Ando03g,Akhmedov03b}.

\subsection{Decaying neutrinos}
\label{sub:Decaying neutrinos}

Non-radiative neutrino decay, which is not satisfactorily constrained,
potentially and significantly changes the SRN flux and spectrum.
Most stringent lower limit on the neutrino lifetime-to-mass ratio comes
from the solar neutrino observations
\cite{Beacom02}--\cite{Eguchi03b} as well as the meson decay experiments
\cite{Barger82}--\cite{Britton94}, which is $\tau/m \gtrsim 10^{-4}$
s/eV.
Since this limit is still very weak, neutrino decay could also change
the detected signal from high-energy astrophysical objects
\cite{Beacom03} or the galactic supernova explosions
\cite{Frieman88}--\cite{Ando04c}, and could alter usual discussions
on the early universe and structure formation \cite{Beacom04b} as well
as on supernova coolings \cite{Fuller88}--\cite{Farzan03}.
In this subsection, we discuss the potential consequence of the neutrino
decay on the SRN flux, with which we can obtain the most stringent limit
on the neutrino lifetime in principle, as pointed out by Ando
\cite{Ando03f}.

Since the flux estimation of SRNs including the neutrino decay is
somewhat complicated, we rather follow a simple and intuitive argument
given in \cite{Ando03f}; exact formulation was recently derived by Fogli
\etal \cite{Fogli04}.
Ando \cite{Ando03f} assumed the following conditions: (i) daughter
neutrinos are active species; (ii) neutrino mass spectrum is
quasi-degenerate ($m_1\approx m_2\approx m_3$); and neutrino mass
hierarchy is normal.
This is because the SRN flux would be most strongly enhanced owing to
the decay and it might give a larger flux than the current observational
limit by SK \cite{Malek03}; in that case the flux limit can be
translated into the limit on the neutrino lifetime.
He also restricted his discussion only to the helicity-conserving mode
of $\bar\nu_3\to \bar\nu_1$ and $\bar\nu_2 \to \bar\nu_1$ for
simplicity, characterized by lifetimes $\tau_3$ and $\tau_2$ of each
mode; in models discussed in \cite{Kim90}--\cite{Giunti92}, the above
condition (ii) strongly suppresses the helicity-flip mode.
In the actual calculation, instead of these lifetimes, he introduced the
``decay redshift'' $z_i^\rmd$ $(i=2,3)$ of the mass eigenstate $\bar\nu_i$
as two free parameters.
If the source redshift $z$ is larger than the decay redshift $z_i^\rmd$,
all the neutrinos $\bar\nu_i$ were assumed to decay, on the other hand
if $z < z_i^\rmd$, completely survive.
The relation between $z_i^\rmd$ and $\tau_i/m$ can be written as
\begin{equation}
\tau_i = \frac{mc^2}{E_\nu} \int_{z_i^\rmd}^0 \frac{\rmd t}{\rmd z}\rmd z
 = \frac{mc^2}{E_\nu H_0}\int_0^{z_i^\rmd}
 \frac{\rmd z}{(1+z)\sqrt{\Omega_{\rm m}(1+z)^3 + \Omega_\Lambda}},
\label{eq:lifetime}
\end{equation}
where we assume $E_\nu = 10$ MeV as a typical neutrino energy when
evaluating the lifetime $\tau_i$ from $z_i^\rmd$.

\begin{figure}[htbp]
\begin{center}
\includegraphics[width=8cm]{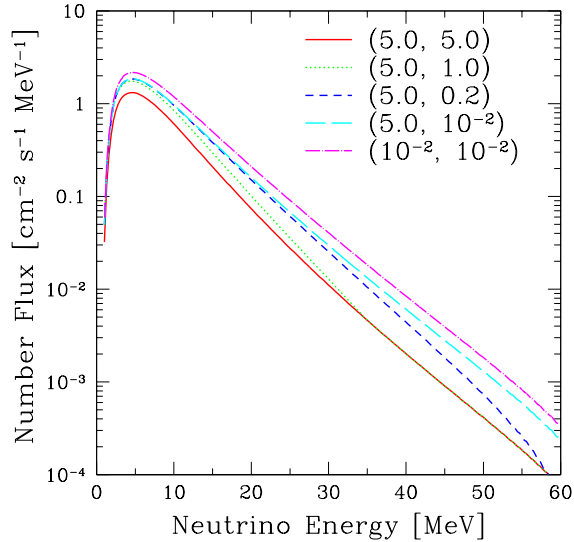}
\vspace{-5mm}
\caption{SRN flux for various parameter sets of decay redshifts,
 calculated with the LL model and $f_\ast = 1$. Each label represents
 ($z_2^\rmd, z_3^\rmd$). This figure is taken from
 \cite{Ando03f}. \label{fig:decay}}
\end{center}
\end{figure}

\Fref{fig:decay} shows the SRN flux for various parameter sets of decay
redshifts $(z_2^\rmd,z_3^\rmd)$ as a function of neutrino energy; the LL
model as the original neutrino spectrum and $f_\ast = 1$ were assumed.
The solid curve in \fref{fig:decay} shows the SRN flux in the case that
$z_2^\rmd = z_3^\rmd = z_{\rm max}(= 5)$, which represents the same flux
as that without the neutrino decay.
Then, the decay of the heaviest mass eigenstate $\bar\nu_3$ was included
by his reducing the value of $z_3^\rmd$, with keeping $\bar\nu_2$
stable.
When $z_3^\rmd=1.0$ (dotted curve), the SRN flux at low energy region
$(E_\nu\lesssim 35~{\rm MeV})$ deviates from the no-decay model.
This is because the neutrinos from supernovae at redshift larger than
$z_3^\rmd=1.0$ are affected by the $\bar\nu_3\to\bar\nu_1$ decay and it
results in the increase of $\bar\nu_\rme$.
Since the neutrino energies are redshifted by a factor of $(1+z)^{-1}$
owing to an expansion of the universe, the decay effect can be seen at
low energy alone.
When the value of $z_3^\rmd$ is reduced to $10^{-2}$, the neutrinos even
from the nearby sources are influenced by the $\bar\nu_3\to\bar\nu_1$
decay, resulting in the deviation over the entire energy range as shown
by the long-dashed curve in \fref{fig:decay}.
If the $\bar\nu_2\to\bar\nu_1$ decay is added, it further enhances the
SRN flux.
In \tref{table:limit}, the SRN flux integrated over the energy range of
$E_\nu>19.3$ MeV is summarized, for the each decay model.
In the second column the lifetime-to-mass ratio is indicated, which
corresponds to each decay redshift, which is obtained using equation
\eref{eq:lifetime}.
The ratio between the predicted flux and the 90\% CL upper limit given
by the SK observation \cite{Malek03} is also shown in the fourth
column.
This result shows that several decaying models may have already been
excluded or severely constrained by the current SRN limit by SK.
It also proves that the SRN observation can potentially give us the
lower limit on the neutrino lifetime as strong as $\sim 10^{10}$
s/eV, which is many orders of magnitude stronger than that by the solar
neutrino observations.
In addition, future detectors such as the HK and UNO (maybe loaded with
Gd) is expected to greatly improve our knowledge of the SRN spectral
shape as well as its flux; if a number of data were actually acquired by
such detectors, the most general and model-independent discussions
concerning the neutrino decay would become accessible.
\Table{\label{table:limit}Predicted SRN flux for various decay
 models. The LL model is assumed as the original neutrino spectrum and
 integrated energy range is $E_\nu >19.3$ MeV. The ratio of the
 prediction and the limit is shown in the fourth column.}
\br
 & ($\tau_2/m,\tau_3/m$) & Predicted flux & &\\
 Model $(z_2^\rmd,z_3^\rmd)$ & (s/eV) & ($f_\ast$ cm$^{-2}$ s$^{-1}$) &
 Prediction/limit\\
\mr
 $(5.0,5.0)$ & $(3.9\times 10^{10},3.9\times 10^{10})$ & 0.43 &
 $0.35 f_\ast$\\
 $(5.0,1.0)$ & $(3.9\times 10^{10},2.4\times 10^{10})$ & 0.55 &
 $0.42 f_\ast$\\
 $(5.0,0.2)$ & $(3.9\times 10^{10},7.7\times 10^9)$ & 0.93 &
 $0.75 f_\ast$\\
 $(5.0,10^{-2})$ & $(3.9\times 10^{10},4.4\times 10^8)$ & 1.0 &
 $0.88 f_\ast$\\
 $(10^{-2},10^{-2})$ & $(4.4\times 10^8,4.4\times 10^8)$ & 1.4 &
 $1.2 f_\ast$\\
\br
\endTable

\section{Conclusions}
\label{sec:Conclusions}

As stressed several times in this paper, we are at an exciting era of
the SRN observation.
This is because the current SK upper limit \cite{Malek03} is {\it just
above} several theoretical predictions using realistic models for the
cosmic SFR and supernova neutrino spectrum, and also because of the
promising technique proposed in \cite{Beacom04a} could greatly improve
performance of the presently working or planned future detectors,
especially for detecting SRNs.
Being at such a situation, we believe that it is timely to review recent
progress of theoretical and observational researches concerning SRNs
from various points of view.
Many physical and astronomical consequences can be derived from the
constraints on the SRNs as we have reviewed through this paper.

Basics of SRN calculation were given in sections \ref{sec:Formulation
and models} and \ref{sec:Flux and event rate of supernova relic
neutrinos}.
Models involved in the SRN calculations are those of the cosmic SFR and
original supernova neutrino spectrum.
Although the SFR-$z$ relation has been intensively studied in recent
years, there remains a fair amount of uncertainty; in fact, the best
estimate of the local SFR density ranges fairly widely as
(0.5--2.9)$\times 10^{-2}h_{70}$ yr$^{-1}$ Mpc$^{-3}$ \cite{Baldry03}.
In order to take this uncertainty into account, we have introduced the
correction factor $f_\ast$ in equation \eref{eq:SFR}, for which a wide
range (0.7--4.2) is still allowed from the SFR observation.
As for the original supernova spectrum, contrary to the traditional
approach using the Fermi-Dirac distribution, we followed the approach
taken by our group \cite{Ando03a,Ando04a}, i.e., adopting the results of
numerical simulations.
Especially following \cite{Ando04a}, we used three neutrino spectra by
different groups, i.e., LL \cite{Totani98a}, TBP \cite{Thompson03} and
KRJ \cite{Keil03}, and performed comparison among models.
Recent progress of neutrino experiments has proven that neutrinos are
massive and mix among different flavours.
This effect was also taken into account appropriately.
As the result, it was found that the uncertainty concerning the original
neutrino spectrum gives difference in the resulting SRN flux by at least
a factor of 2--4.
The expected event rate at a water \v Cerenkov detector with a fiducial
volume of 22.5 kton was found to be 0.25--1 $f_\ast$ yr$^{-1}$ (for
$E_\rme > 19.3$ MeV) and 1--2 $f_\ast$ yr$^{-1}$ (for $E_\rme > 10$
MeV), depending on these supernova models (see \tref{table:event
rate}).

Besides the flux estimation, a careful discussion about background
events is definitely necessary in order to investigate the
detectability.
This has been thoroughly argued in \cite{Ando03a} and we followed them
in \sref{sec:Detectability and observational upper limit}.
The most serious background comes from decay products of invisible
muons, which are produced almost at rest inside the detector and that is
why ``invisible.''
Because of the background, the detection is severely restricted; for a
pure-water \v Cerenkov detector with the size of SK, it would take about
(maybe more than) 10 years in order to reach the signal-to-noise ratio
of a few (see \fref{fig:SN_ratio}).
We also discussed performance of other detectors such as SNO and KamLAND
in the same section, by mainly following \cite{Ando03a,Strigari04}.
The current observational upper limit by SK \cite{Malek03} and the
proposed technique that potentially improves the performance of water \v
Cerenkov detectors \cite{Beacom04a} were briefly introduced there.
In addition, we showed that the SK limit on the SRN flux can be fairly
well reproduced by a simple statistical argument using equation
\eref{eq:SN ratio}.

\Sref{sec:Implication for cosmic star formation history} was devoted to
the discussion on the current status and future prospects for obtaining
limits on the cosmic SFR from the SRN observations.
We first showed that the current SRN limit from SK \cite{Malek03}
already sets a stringent limit on the local SFR value, as discussed in
\cite{Fukugita03}.
Surprisingly, it might rule out a part of local SFR value allowed by the
current observations using the light as shown in
\tref{table:constraints}.
Performance of future detectors loaded with Gd as proposed in
\cite{Beacom04a} was investigated in detail using the MC simulation,
following discussion in \cite{Ando04a}.
The distribution of the best-fit values of two parameters concerning SFR
\eref{eq:parameterization} was obtained by the generated $10^3$ MC data
and by accompanying likelihood analyses.
If one of two free parameters was fixed somehow in advance using other
observations, then the SRN observation can constrain the remaining
parameter by accuracy of $\sim 0.3 (f_\ast V_{\rm eff}/22.5 ~{\rm
kton} \cdot 5 ~{\rm yr})^{-1/2}$; this number would soon become quite
significant if the future mega-ton class detectors such as HK or UNO
loaded with Gd started data-taking.

Finally in \sref{sec:Implication for neutrino properties}, SRN
constraints on particle physics models of neutrinos were discussed.
Neutrino oscillations in the case of inverted mass hierarchy
\cite{Ando03c}, resonant spin-flavour conversions induced by the
neutrino magnetic moment (see \cite{Ando03g} and references therein) and
neutrino decay \cite{Ando03f,Fogli04} are expected to considerably
change the SRN signal at the detectors.
In particular for neutrino decay model, the SRN observation potentially
gives the most stringent constraint on the neutrino lifetime, compared
with a galactic supernova \cite{Frieman88}--\cite{Ando04c},
high-energy astrophysical objects \cite{Beacom03} as well as the most
reliable (but weak) limit at present by the solar neutrino observations
\cite{Beacom02}--\cite{Eguchi03b}.

\ack
SA's work is supported by a Grant-in-Aid for JSPS Fellows.
KS's work is supported in part by a Grant-in-Aid for Scientific
Research provided by the Ministry of Education, Culture, Sports, Science
and Technology of Japan through Research Grant No S14102004,
a Grant-in-Aid for Scientific Research on priority Areas
No 14079202.

\section*{References}

\end{document}